\documentclass[conference]{IEEEtran}
\IEEEoverridecommandlockouts

\usepackage{cite}
\usepackage{amsmath,amssymb,amsfonts}
\usepackage{algorithm}
\usepackage{algorithmic}
\usepackage{graphicx}
\usepackage{textcomp}
\usepackage{xcolor}

\usepackage{subcaption}
\usepackage{amsthm}
\usepackage{tabularx}
\usepackage{booktabs}
\usepackage{multirow}
\def\BibTeX{{\rm B\kern-.05em{\sc i\kern-.025em b}\kern-.08em
    T\kern-.1667em\lower.7ex\hbox{E}\kern-.125emX}}
\begin{document}

\title{FBChain: A Blockchain-based Federated Learning  Model with Efficiency and Secure Communication
}


\author{
	\IEEEauthorblockN{
		Yang Li\IEEEauthorrefmark{10}\IEEEauthorrefmark{2}, 
		Chunhe Xia \IEEEauthorrefmark{2}\IEEEauthorrefmark{3},
		and Tianbo Wang\IEEEauthorrefmark{9}\IEEEauthorrefmark{1}}
	\IEEEauthorblockA{\IEEEauthorrefmark{10}School of Computer Science and Engineering,  Beihang University, Beijing, China }
	\IEEEauthorblockA{\IEEEauthorrefmark{2}Key Laboratory of Beijing Network Technology, Beihang University, Beijing, China}
	\IEEEauthorblockA{\IEEEauthorrefmark{3}Guangxi Collaborative Innovation Center of Multi-Source Information  Integration and Intelligent Processing,\\
		Guangxi Normal University, Guilin,China} 
	\IEEEauthorblockA{\IEEEauthorrefmark{9}School of Cyber Science and Technology, Beihang University,  Beijing,China} 
	\IEEEauthorblockA{\{johnli,xch,wangtb\}@buaa.com,corresponding author: wangtb@buaa.com}
} 

\maketitle

\begin{abstract}
Privacy and security in the parameter transmission process of federated learning are currently among the most prominent concerns. However, there are two thorny problems caused by unprotected communication methods: ``parameter-leakage'' and ``inefficient-communication". 
This article proposes  \underline{B}lock\underline{chain}-based \underline{F}ederated Learning (FBChain) model for federated learning parameter communication to overcome the above two problems. First,  we utilize the immutability of blockchain to store the global model and hash value of local model parameters in case of tampering during the communication process, protect data privacy by encrypting parameters, and verify data consistency by comparing the hash values of local parameters, thus addressing the ``parameter-leakage" problem. Second, the  Proof of Weighted Link Speed (PoWLS) consensus algorithm comprehensively selects nodes with the higher weighted link speed to aggregate global model and package blocks, thereby solving the ``inefficient-communication" problem. Experimental results demonstrate the effectiveness of our proposed FBChain model and its ability to improve model communication efficiency in federated learning.
\end{abstract}


\section{Introduction}

With the increasing privacy awareness of people and the enactment of relevant privacy laws, federated learning (FL) is emerging as a viable solution to train machine learning models with decentralized datasets while protecting privacy  \cite{Kairouz2019AdvancesAO}. In vanilla FL, clients train local model utilizing local dataset. Then, they communicate with a parameter server, accept all the updated local models to aggregate global model each round until model converges. However, FL is a double-edged sword  \cite{Mothukuri2021ASO, RodriguezBarroso2022SurveyOF}. On the positive side,  FL protects training dataset privacy and security, a large amount of data exists in isolated silos, and FL enables their use for training models in a safe environment. On the negative side, because the training dataset is decentralized and the global model is aggregated by different local models, the communication of local model between clients and server is frequent, but the communication efficiency and security cannot be guaranteed  \cite{Li2019FederatedLC}. Given the local model parameter, research on communication security and efficiency focuses on model encryption and lower communication rounds.  Especially, our research domain helps to protect local model parameters against malicious attacks that may result in data tampering or leakage, while concurrently enhancing communication efficiency.

There are two inevitable problems in federated learning communication: ``parameter-leakage" and ``inefficient-communication"  \cite{Nguyen2021FederatedLF, Alazab2022FederatedLF,  Wahab2021FederatedML,Huang2023StochasticCA}. The local model parameters transmission from clients and server via network,  thus getting aggregated global model. Nevertheless, the network may be under malicious attack, and the link speed of different clients and server may be difference. ``Parameter-leakage" refers to the local model parameter being tampered or leakage because of unsafety communication method, attackers may get private data from leaked parameters. ``Inefficient-communication" refers to the parameter transmission speed that may be slow. These two problems indicate it is thorny for the FL to achieve efficient and secure communication. 


Benefiting from the great success of FL in privacy protection, most researchers consider privacy during FL communication and use compression or combine it with blockchain to improve security and efficiency. Compression local model parameters can lower transmission data size, according to compress origin data to low-rank or a random mask  \cite{ Xu2020TernaryCF, Sun2020AdaptiveFL}. Blockchain can provide a reliable way to transmit combination with encryption mechanism, store local model and global model parameters in blocks to ensure the security of model parameters  \cite{Cui2022CREATBC, Issa2022BlockchainBasedFL,Qu2022BlockchainenabledFL}. Unfortunately, these two mainstreams are inadequate to address the problem of FL communication efficiency and security. First, the compression of parameters has differences from origin local model, makes it challenging to perceive better performance on model training  \cite{Asad2023LimitationsAF,Haddadpour2020FederatedLW}. Second, the blockchain combined method has limitations on block size, and every block will be stored in each node, which will occupy a significant amount of storage space  \cite{Huang2023BlockchainBasedFL,Zhu2022BlockchainempoweredFL}.


To tackle the obstacles above, we propose the blockchain-based federated learning (FBChain) model, which consists of two main components: the model and Proof of Weighted Link Speed consensus algorithm(PoWLS) in the blockchain.
The former integrates asymmetric encryption, symmetric encryption and hash computation, thus solving the problem of ``parameter-leakage". It considers the parameter tampered with or leaked during transmission, while reducing the amount of data that needs to be stored in the block, trying to ensure data privacy and security while reducing the pressure on blockchain storage. The latter involves selecting a group of nodes with strong communication capabilities to aggregate global model and package blocks, thereby addressing the  ``inefficient-communication" problem. Experimental results demonstrate that our model can not only have the effectiveness on machine learning model training, but also make different communication efficiency during parameter communication.

Our contributions can be summarized as follows:
\begin{itemize}
	\item We propose the FBChain model by storing the global model and the hash values of local model parameters in the blockchain, ensures the immutability of the global model and reduces the amount of data stored in the blockchain. Local model parameters are processed by encryption, and the aggregation node compares the received parameters with the saved hash value in the blockchain, achieving consistency verification of the local model and enhancing the security of the data communication process.
	\item We propose a PoS and DPoS inspired consensus algorithm. By comprehensively considering the link speed of nodes in the blockchain network, a group of nodes with high link speed and low latency is selected to take turns aggregate global parameters and package blocks, improve  communication efficiency.
	\item Experiments on real-world datasets demonstrate that our model outperforms baseline approaches in terms of communication efficiency.
\end{itemize}
The rest of this article is organized as follows. We first give a comprehensive review of related works in Section \ref{RELATED WORK}. Next, we demonstrate the background of this article in Section \ref{BACKGROUND},
followed by the presentation of  detailed design of FBChain in Section \ref{Blockchain-based Federated Learning Model}.  After that, we conduct a series of experiments on two public datasets to evaluate FBChain in Section \ref{PERFORMANCE EVALUATION}. Finally,
Section  \ref{Conclusion} concludes this work and discusses future directions.

\section{RELATED WORK}
\label{RELATED WORK}

In this section, we briefly introduce some related works about federated learning improvement on transaction efficiency and security. 
\subsection{Improvement on communication efficiency}

Although federated learning can build a global model without sharing training data, a large number of model parameters need to be exchanged during the construction process. Jakub Konecný et al.  \cite{Konecn2016FederatedLS} proposed two ways to reduce the uplink communication costs: structured updates and sketched updates. Yunlong Lu et al.  \cite{Lu2021CommunicationEfficientFL} proposed a blockchain-empowered federated learning scheme in digital twin edge networks to strengthen communication security and data privacy protection. Su Liu et al. \cite{Liu2021FedCPFAE} proposed an efficient-communication approach, which consists of three parts, provides a customized local training strategy for vehicular clients to achieve convergence quickly through a constraint item within fewer communication rounds. K. Li et al.  \cite{Li2022CBFLAC} proposed a coreset-based FL (CBFL) framework to improve communication efficiency in federated learning. Instead of training model on full datasets with a regular network model, CBFL uses a much smaller well-matched evolutionary network model on coreset. Qing Han et al.  \cite{Han2022PCFedPA} proposed PCFed, a novel privacy-enhanced and communication-efficient FL framework to provide higher model accuracy with rigorous privacy guarantees and great communication efficiency. Wei Liu et al. \cite{Liu2021DecentralizedFL} proposed a general DFL framework, which implements both multiple local updates and multiple inter-node communications periodically, to strike a balance between communication efficiency and model consensus.
\subsection{Improvement on communication security}
During the communication of parameters of the federal learning model, the communication may under malicious attack, and the model parameters may be tampered with or leaked.  Jiaqi Zhao et al.  \cite{Zhao2022PVDFLAP}  proposed a privacy protection and verifiable decentralized co-learning framework called PVD-FL, which can realize secure deep learning model training under a decentralized architecture. Zhe Peng et al.  \cite{Peng2021VFChainEV} proposed VFChain, a verifiable and auditable joint learning framework based on the blockchain system. Yuanhang Qi et al.  \cite{Qi2021PrivacypreservingBF} proposed a blockchain-based joint learning framework to realize decentralized, reliable, and safe joint learning without a centralized model coordinator. 

During the transmission of parameters of the federal learning model, the data may be tampered with or leaked.Jiaqi Zhao et al.  \cite{Zhao2022PVDFLAP}  proposed a privacy protection and verifiable decentralized co-learning framework called PVD-FL, which can realize secure deep learning model training under a decentralized architecture.Zhe Peng et al.  \cite{Peng2021VFChainEV} proposed VFChain, a verifiable and auditable joint learning framework based on the blockchain system. Yuanhang Qi et al.  \cite{Qi2021PrivacypreservingBF} proposed a blockchain-based joint learning framework to realize decentralized, reliable and safe joint learning without a centralized model coordinator. Jungjae Lee et al.  \cite{Lee2022DAGBasedBS}  proposed a layered blockchain system, using public blockchain for a joint learning process without a trustworthy curator. This can prevent model poisoning attacks and provide security updates for the global model. Xiaoyuan Liu et al.  \cite{Liu2021PrivacyEnhancedFL}  proposed a privacy-enhanced FL (PEFL) framework, which uses homomorphic encryption as the basic technology and provides a channel for the server to punish the poisoned by extracting the effective gradient data of the number function.

\section{BACKGROUND}
\label{BACKGROUND}
\subsection{Symmetric encryption and asymmetric encryption}

In cryptography, encryption methods can be divided into two categories of cryptographic algorithms based on the characteristics of the key: symmetric encryption and asymmetric encryption. Symmetric encryption algorithms are a type of encryption algorithm in cryptography. In this type of algorithm, the same key is used for both encryption and decryption, or two keys that can be easily calculated from each other are used. In symmetric encryption algorithms, because the same key is used for encryption and decryption, both parties in communication must jointly select and keep the same key. Each party must ensure the privacy and security of the symmetric key in order to achieve the confidentiality and integrity of the data. Symmetric encryption has the advantages of fast speed and simple algorithm, but it also has the disadvantages of complex key distribution and management, high cost, and inability to be used for digital signatures.
Asymmetric encryption algorithms contain two different keys, the public key $k_p$ and the private key $k_s$. $k_p$ can be made public to other nodes, while $k_s$ can only be kept by oneself. It is easy to calculate $k_p$ from $k_s$, but it is difficult to calculate $k_s$ from $k_p$. Asymmetric encryption algorithms have the following characteristics:
\begin{itemize}
	\item Encryption and decryption are performed respectively by $k_p$ and $k_s$, $k_p \neq k_s$.
	
	Asymmetric encryption: $X \rightarrow Y : Y = Enc(X, k_p)
	$.
	
	Asymmetric decryption: $Y \rightarrow X: X =  Dec(Y, k_s) = D(E(X, k_p),k_s)$.

	Hence, X is the content to encrypt, 	Y is the ciphertext, $Enc$ is the encryption function, $Dec$ is the decryption function.
	\item Can't get $k_s$ from $Enc$ and $k_p$.
	\item Both $k_p$ and $k_s$ can be used as encryption key and decryption key. 
	$Y \rightarrow X: X = D(E(X, k_p),k_s) = E(D(X,k_s),k_p)$
\end{itemize}

The advantages of asymmetric encryption algorithms are that key distribution and management are simple, and it is relatively easy to implement digital signatures and key exchange. The disadvantage is that the algorithm is more complex and the encryption and decryption speed is slower.

\subsection{PoS Consensus algorithm}

One of the fundamental problems in distributed systems is how to ensure that the data of all nodes in a distributed system cluster are completely identical and can reach consensus on a proposal. Consensus algorithms focus on studying the process of distributed nodes reaching consensus. How to make all nodes in a distributed system cluster reach consensus in a complex, open, and untrusted Internet environment is still one of the challenges in the field of distributed computing.

The Proof of Stake (PoS) consensus algorithm using stake as witness nodes selection criteria, nodes with the highest stake, rather than the highest computing power, are awarded the right to record transactions. The stake is reflected in the node's ownership of a specific amount of currency, called Coin days. The PoS consensus algorithm has advantages such as high efficiency and low resource consumption.

\section{Blockchain-based Federated Learning Model}
\label{Blockchain-based Federated Learning Model}
To address the issue of man-in-the-middle attacks and improve parameter communication efficiency during the training process of federated learning models, this article proposes the FBChain model based on federated learning and blockchain. FBChain is based on a blockchain network, 
assume there have $\rho$ nodes in the blockchain network. The model is defined as follows:

$$FBChain=\{PA,LT,BP,P_L^{\Delta,e},P_G^{\Gamma,r},PoWLS,CR\}$$

Hence, $PA=\{PA_1, PA_2,\cdots, PA_\alpha\}, 0 \leq \alpha \leq \rho \cap \alpha \in \mathbb{Z}^+$, represents the set of global model aggregation packaging nodes, where $\alpha$ represents $PA$ number and $\mathbb{Z}^+$ represents the set of positive integers. The $PA$ nodes are responsible for aggregating the local model parameters ($LM$) to global model ($GM$) in federated learning and packaging transactions into blockchain.

$LT=\{LT_1,LT_2,\cdots,LT_{\beta}\}, 0 \leq \beta \leq \rho \cap \beta \in \mathbb{Z}^+$, represents the set of local training nodes, where $\beta$ represents the $LT$ nodes number. The $LT$ nodes have their own data sets.

$BP=\{BP_1,BP_2,\cdots, BP_\gamma\}, 0 \leq \gamma \leq \rho \cap  \gamma\in \mathbb{Z}^+$, represents the blockchain propagation nodes, where $\gamma$ represents $BP$ nodes number. The blockchain propagation nodes do not participate in the federated learning training process but only propagation blocks.

$P_L^{\Delta,e}=\{P_L^{1,e},P_L^{2,e},\cdots, P_L^{\beta,e}\}, 1 \leq \Delta \leq \beta$, is local model set, where $\Delta$ represents the node number, $e$ represents the local parameter update round, $P_L^{\Delta,e}$ is $LM$ generated by the $LT_{\Delta}$ in round $e$, and will be sent to $PA$ for aggregation. 

$P_G^{\Gamma,r}=\{P_{G}^{1,r}, P_{G}^{2,r}, \cdots, P_{G}^{\beta,r}\}, 1 \leq \Gamma \leq \beta$,  is global model set, where $G$ represents $GM$, where $\Gamma$ represents the $PA$ who aggregated the global model, and $r$ represents the global parameter update round. The unified $GM$ is obtained by aggregating local parameters.

$PoWLS$ represents Proof of Weighted Link Speed consensus algorithm, which comprehensively consider nodes' link speed and transmission delay in the blockchain network to obtain a weighted value, as the basis for selecting nodes to be $PA$, improve the communication efficiency in the FBChain.

$CR=\{CR_1,CR_2,\ldots,CR_{\zeta}\},\zeta\in \mathbb{Z}^+$ represents nodes' credit score, where $\zeta$ represents the node number. In the model, the credit score evaluates the node's performance in the federated learning training process. The higher the credit score, the better the node's local model performance in the global model aggregation process. 

\begin{figure*}[htbp]
\centering
\includegraphics[width=\textwidth]{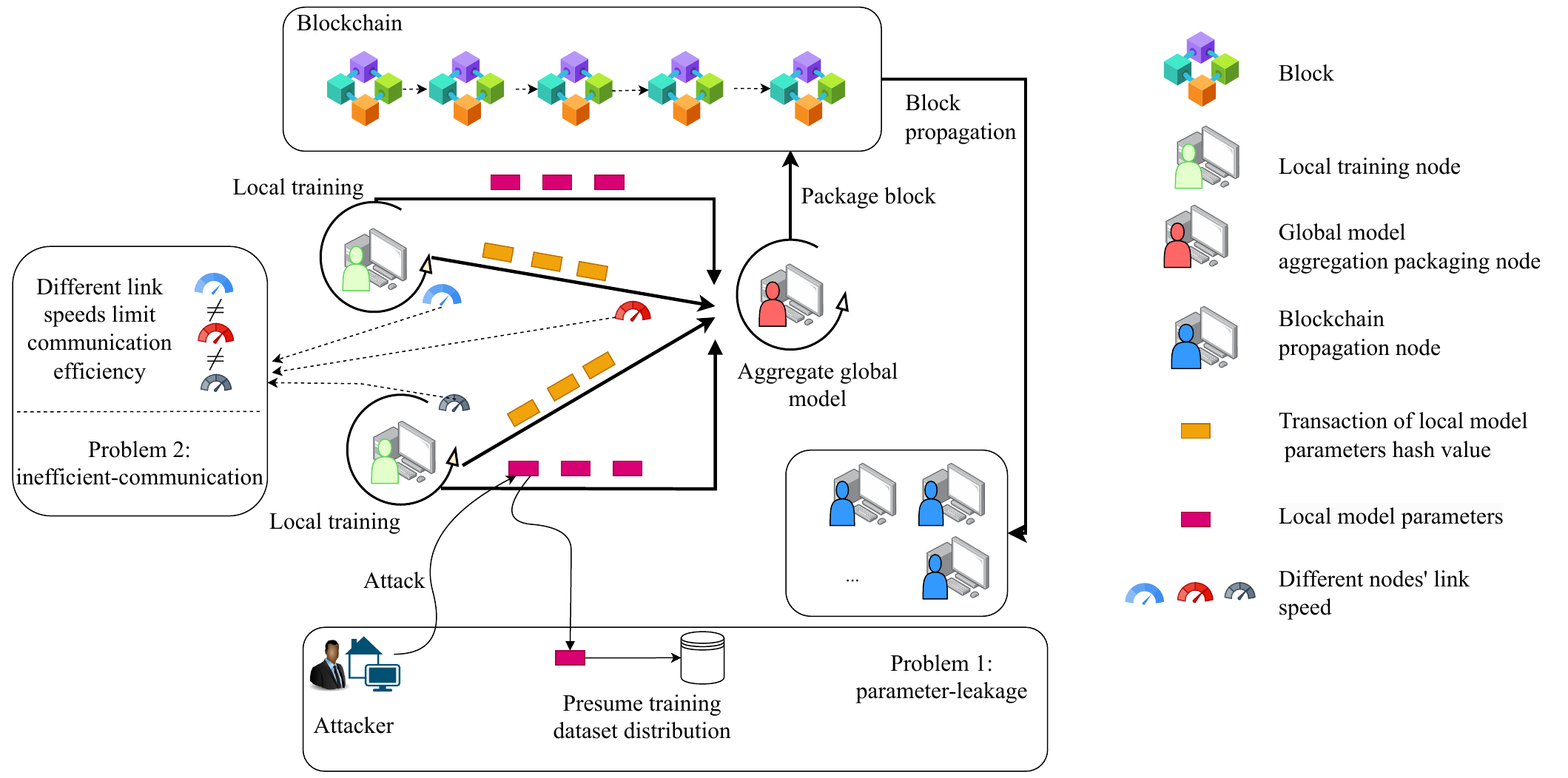}
\caption{FBChain Model Architecture}
\label{FBChain Model Architecture}
\end{figure*}

The model architecture diagram is shown in Figure \ref{FBChain Model Architecture}.


\begin{table}[h]
\centering
\caption{List of Notations}
\label{tab:List of Notations}
\begin{tabularx}{\columnwidth}{lX}
	\toprule
	Notations & Descriptions  \\
	\midrule
	$\rho$ & total nodes number \\
	$\mathbb{B}$ & block in blockchain \\
	$hash$ &  the hash value  \\
	$\mathbb{P}_{L,u}^{\delta,e}$ & the update local model in $e$ round of local training node $\delta$  after training\\
	$\mathbb{T}(hash_{\mathbb{P}_{L,u}^{\delta,e}})$ &  transaction contains $\mathbb{P}_{L,u}^{\delta,e}$ hash value  \\
	$\mathbb{S}_{P}$ &  serialized data of $\mathbb{P}_{L,u}^{\delta,e}$  \\
	$\mathbb{C}(\mathbb{S}_{P})$ &  compressed data of  $\mathbb{S}_{P}$ \\
	$\mathfrak{S}|K_{LT_{\delta}}^e$ &  $LT_{\delta}$'s symmetric encryption key in round $e$  \\
	$\mathfrak{S}_{\mathbb{C}}^{P}$ &  symmetric encrypted compressed $LT_{\delta}$'s update local model  \\
	$\mathbb{SE}$  &  symmetric encryption algorithm  \\
	$\mathbb{PK}_{\mathfrak{P}}^{\iota}$ &  $\iota$th $PA$'s public key in package nodes list of current round  \\
	$\mathbb{SK}_{\mathfrak{P}}^{\iota}$ &  $\iota$th $PA$'s private key in package nodes list of current round   \\     
	$\mathbb{AE}$ & asymmetric encryption algorithm   \\    
	$AEC$ & asymmetrically encrypted symmetric encryption key \\
	$CR$ & Credit score \\
	\bottomrule
	
\end{tabularx}
\end{table}

\subsection{Model process}
In the FBChain model, we use blockchain to store $GM$, and the hash values of $LM$, the $r$ round block is $\mathbb{\mathbb{B}}_{r}$, $LT$ get $\mathbb{\mathbb{B}}_{r}$ to continue next step training. During model aggregation, $PA$ receives $P_L^{\delta,e}$ directly from $LT_{\delta}$. Before transmitting $P_L^{\delta,e}$ to $PA$, $LT_{\delta}$ stores its hash value on the blockchain and compresses $P_L^{\delta,e}$ into a compressed file. This file is then encrypted symmetrically, and the encryption key is encrypted asymmetrically using $PA$'s public key to prevent tampering during transmission. Once $PA$ receives the encrypted data, it decrypts the symmetric encryption key using its private key and then decrypts the data itself. The resulting local model compared to the hash value stored on the blockchain to ensure that the model wasn't tampered with. If the hash values match, it means that $P_L^{\delta,e}$ was transmitted without tampering, ensuring consistency and tamper resistance during data transmission.
We introduce the credit score in FBChain, where nodes with poor local model training results will have their credit scores deducted, and credit scores less than a threshold will limit how often the node participates in global model aggregation.

The process of the FBChain model is as follows:

\textbf{Initialize Local Model. }
 
Utilize the $PoWLS$ consensus algorithm to choose $\alpha$ nodes, excluding the $LT$ node, from the blockchain network, considering their weighted link speeds. These selected nodes form the package nodes list ($\mathfrak{P}$). $LT_{\delta}$, $\delta \in [1,\beta]$ in blockchain network will initialize $LM$ by model weight random generation.

\textbf{Process Updated Local Model.}

After $\eta$ epochs of local training, $LT_{\delta}$ obtains a locally updated model, denoted as $\mathbb{P}_{L,u}^{\delta,e}$. $LT_{\delta}$ then calculates the hash value of $\mathbb{P}_{L,u}^{\delta,e}$, denoted as $hash_{\mathbb{P}_{L,u}^{\delta,e}}$, and adds it to a transaction, denoted as $\mathbb{T}(hash_{\mathbb{P}_{L,u}^{\delta,e}})$. The transaction is broadcast on the blockchain network.

$LT_{\delta}$ serializes and transforms $\mathbb{P}_{L,u}^{\delta,e}$ into a serialized data format, denoted as $\mathbb{S}_{P}$, and compresses it to reduce the communication data size. The compressed serialized data is denoted as $\mathbb{C}(\mathbb{S}_{P})$. $LT_{\delta}$ then initializes a symmetric encryption key, denoted as $\mathfrak{S}|K_{LT_{\delta}}^e$, which is $LT_{\delta}$'s symmetric encryption key in round $e$. $LT_{\delta}$ encrypts $\mathbb{C}(\mathbb{S}_{P})$ with $\mathfrak{S}|K_{LT_{\delta}}^e$ using a symmetric encryption algorithm, denoted as $\mathbb{SE}$, to obtain a symmetrically encrypted compressed model, denoted as $\mathfrak{S}_{\mathbb{C}}^{\mathbb{P}_{L,u}^{\delta,e}}$.

As $\mathfrak{S}|K_{LT_{\delta}}^e$ is important, $LT_{\delta}$ performs asymmetric encryption on it using $\mathfrak{P}_{\iota}$'s asymmetric encryption public key, denoted as $\mathbb{PK}_{\mathfrak{P}}^{\iota}$. The asymmetrically encrypted symmetric encryption key is denoted as $AEC_{\mathfrak{S}|K}^{LT_{\delta}^e} = \mathbb{AE}(\mathfrak{S}|K_{LT_{\delta}^e}, \mathbb{PK}_{\mathfrak{P}}^{\iota})$, where $\mathbb{AE}$ is an asymmetric encryption algorithm.

By storing only the hash value of local models on the blockchain, we can reduce the storage space and block size required. This approach can help reduce block propagation delays and ensure the integrity and confidentiality of local models during transmission.

\textbf{Local Model transmission. }

In FBChain, we use the credit score $CR$ to assess the performance of nodes, and set a threshold $CR_{TH}$ for local model transmission from nodes to $PA$, limiting nodes with poor $CR$ communication time. This helps to reduce the communication of models with poor performance and improve communication efficiency.

Before transmitting the local updated model $\mathbb{P}_{L,u}^{\delta,e}$ to $\mathfrak{P}_{\iota}$ in round $e$, FBChain checks $LT_{\delta}$'s $CR$, denoted as $CR_{\delta}$. If $CR_{\delta}$ meets the threshold requirement ($CR_{\delta} \geq CR_{TH}$), $LT_{\delta}$ can transmit $P_L^{\delta,e}$ to $\mathfrak{P}_{\iota}$ without limitation. Otherwise, if $CR_{\delta}$ is lower than $CR_{TH}$, $LT_{\delta}$ is limited to transmitting $P_L^{\delta,e}$ to $\mathfrak{P}_{\iota}$ only once every $\kappa$ rounds.

Assuming that there are $\lambda$ nodes ($LT$) that can transmit to $\mathfrak{P}_{\iota}$, $LT_{\mu}$, where $\mu \in [0,\lambda]$, transmits an asymmetrically encrypted symmetric encryption key , $AEC_{\mathfrak{S}|K}^{LT_{\mu}^e}$, a compressed and symmetrically encrypted local updated model, $\mathfrak{S}_{\mathbb{C}}^{\mathbb{P}_{L,u}^{\mu,e}}$, and a nonce of symmetric encryption,  $\mathbb{SE}_{LT_{\mu}}^{nonce}$, which is a unique random number used during symmetric encryption.

$\mathfrak{P}_{\iota}$ performs asymmetric decryption on $AEC_{\mathfrak{S}|K}^{LT_{\mu}^e}$ using its private key $\mathbb{SK}_{\mathfrak{P}}^{\iota}$ to obtain the symmetric encryption key of $LT_{\mu}$ in round $e$ . $\mathfrak{P}_{\iota}$ then performs symmetric decryption on $\mathfrak{S}_{\mathbb{C}}^{\mathbb{P}_{L,u}^{\mu,e}}$ using $\mathfrak{S}|K_{LT_{\mu}^e}$ and $\mathbb{SE}_{LT_{\mu}}^{nonce}$ to obtain the compressed serialized $\mathbb{P}_{L,u}^{\mu,e}$.

After decompressing $\mathbb{C}(\mathbb{S}_{\mathbb{P}_{L,u}^{\mu,e}})$, we obtain the serialized $\mathbb{P}_{L,u}^{\mu,e}$,
, which can be loaded to obtain the original $\mathbb{P}_{L,u}^{\mu,e}$.

\textbf{Local Model Verify and Global Model Aggregate. }

After receiving the locally updated model $\mathbb{P}_{L,u}^{\mu,e}$ from $LT_{\mu}$, $\mathfrak{P}_{\iota}$ checks if its hash value equals $hash_{\mathbb{P}_{L,u}^{\delta,e}}$ in $\mathbb{T}(hash_{\mathbb{P}_{L,u}^{\delta,e}})(\mu = \delta)$. If the hash value matches, it indicates that the model has not been tampered with during transmission.

Using the untampered $\mathbb{P}_{L,u}^{\mu,e}$, $\mathfrak{P}_{\iota}$ performs an accuracy verification on a self-test dataset. If the test accuracy in $\mathfrak{P}_{\iota}$ is greater than $P_G^{\epsilon, {e-1}}$ or within a certain threshold $t_{acc}$, $\mathbb{P}_{L,u}^{\mu,e}$ can be added to the available local model update group, $ALMG$. Otherwise, if the test accuracy of $\mathbb{P}_{L,u}^{\mu,e}$ in $\mathfrak{P}_{\iota}$ is lower than the accuracy of $P_G^{\epsilon, {e-1}}$ minus $t_{acc}$, $\mathbb{P}_{L,u}^{\mu,e}$ is added to the unavailable local model update group, $ULMG$.

Finally, $\mathfrak{P}_{\iota}$ aggregates the locally updated models received from $LT$ to obtain the global model $P_G^{\iota, {e}} = \sum_{\omega = 1}^{\beta} (P_{L,u}^{\omega,e}) /{\beta}$.

\textbf{Block Package. } 

After aggregating the local updated models and verifying their accuracy, $\mathfrak{P}_{\iota}$ packages the transaction of local model hash value  $\mathbb{T}(hash_{\mathbb{P}_{L,u}^{\delta,e}})$ and the aggregated global model $P_G^{\iota, {e}}$ into a block $\mathbb{B}$, which is then broadcasted to $LT$ and $BP$. $LT$ retrieves $P_G^{\iota, e}$ from $\mathbb{B}$, and performs the next round of local updates based on $P_G^{\iota, e}$ until the training round limit is reached or the expected results are achieved.

\subsection{Proof of Weighted Link Speed Consensus Algorithm}

In the FBChain model, we introduce a consensus algorithm called Proof of Weighted Link Speed (PoWLS). PoWLS takes into account the weighted value of nodes when selecting package nodes. For each node $Node_{\psi}, 0 \leq \psi \leq \rho$, we calculate weighted value, $WV$, based on the node's link speed, $D_{\psi}$, and transmission delay, $TD_{\psi}$. Nodes with higher $WV$ are more likely to be selected as $PA$. By comprehensively considering the network conditions of nodes and selecting nodes with better network conditions and high transmission efficiency, PoWLS improves the efficiency of parameter network transmission in federated learning.

The consensus algorithm process is as follows:

\textbf{Weighted Link Speed Calculate. } Calculate the  $WV$ of  $Node_{\psi}$ based on Equation  \ref{equ:WV} in the blockchain network except for the local training node, and sort them in descending order. 

%

%
%
%
%
\begin{equation}
\begin{aligned}
	WV_{\psi} =  \upsilon \times D_{\psi} + \phi \times (1/TD_{\psi})
\end{aligned}
\label{equ:WV}
\end{equation}
Among them, $\upsilon$, $\phi$ represent the weights of $Node_{\psi}$'s $D, TD$ respectively.

\textbf{Choose Global Model Aggregation Packaging Nodes. } Select the top $\tau$ nodes with high $WV$ to enter $\mathfrak{P}$, and nodes in $\mathfrak{P}$ are $PA$. $PA$ will broadcast transactions received between $\mathfrak{P}$, and aggregate global model separately, the $PA$ with highest $WV$ will add its packaged block into blockchain. 

$$
\begin{aligned}
WV_1 \geq WV_2 \geq \ldots \geq WV_{\tau-1} \\
\geq WV_{\tau} \geq WV_{\tau+1} \geq \ldots \geq WV_{\eta}
\end{aligned}
$$
If the $WV$ of the $\tau$th and $(\tau+1)$th nodes are equal, and only the first $\tau$ nodes are selected to enter $\mathfrak{P}$, the nodes are chosen to join the packaging queue in order of priority based on their  $D$, $TD$.


\textbf{Package Blocks.} For the local update models from $LT$, the $PA$ in $\mathfrak{P}$ take turns aggregating these models. After aggregation, $GM$ will be packaged into a transaction and added to the block with other transactions in the blockchain network during this period.


\textbf{Credit Score and Token Reward.} Nodes in the blockchain network receive $CR$ and token rewards based on their performance. $LT$ nodes in the active local model group (ALMG) are rewarded with $CR^r$ while $LT$ nodes in the unselected local model group (ULMG) are punished with $CR^p$, where $CR^r$ and $CR^p$ are positive and negative real numbers, respectively. Token rewards, denoted as $TR$, are distributed to nodes based on the contribution of their local model to the global model. The total token reward for each round of federated learning is fixed and $LT_{\psi}$ splits it with other $LT$ nodes. If $P_L^{\psi,e}$ performs better than other local models, $LT_{\psi}$ receives a larger share of the token reward, denoted as $TR_{\psi}$, which is calculated using Equation \ref{equ:TR}.
\begin{equation}
\begin{aligned}
	&TR_{\psi} = \\
	&(EX_{\psi}+ t_{acc})/\sum_{i = 1}^{\beta}((EX_{i} + t_{acc})) 
	* TR_{total} 
\end{aligned}
\label{equ:TR}
\end{equation} 

Hence, $EX_{\psi}$ is the $LT_{\psi}$'s $LM$ accuracy difference with the value of the previous round's global model in the global test dataset.

\section{PERFORMANCE EVALUATION}
\label{PERFORMANCE EVALUATION}
All the experiments were conducted on a virtual machine with one  NVIDIA V100 GPU, two Intel Golden 6240 CPUs and 131.43 GB of RAM. All experiments involved 20 devices for FBChain, Vanilla Federated Average model, and VBFL. Each device in FBChain, vanilla federated average learning model, and VBFL adopted $FedAvg$ and $MNIST\_CNN$  \cite{McMahan2016CommunicationEfficientLO} network structure, the training sets are randomly assigned to different parts of the same size, with 5 local training epochs every training round, the learning rate is set 0.01, batch size 10. In PoWLS $\upsilon$ is set 1, $\phi$ is set 100.

\subsection{Effectiveness of FBChain}

\begin{figure}[htbp]
\centering
\includegraphics[width=.5\textwidth]{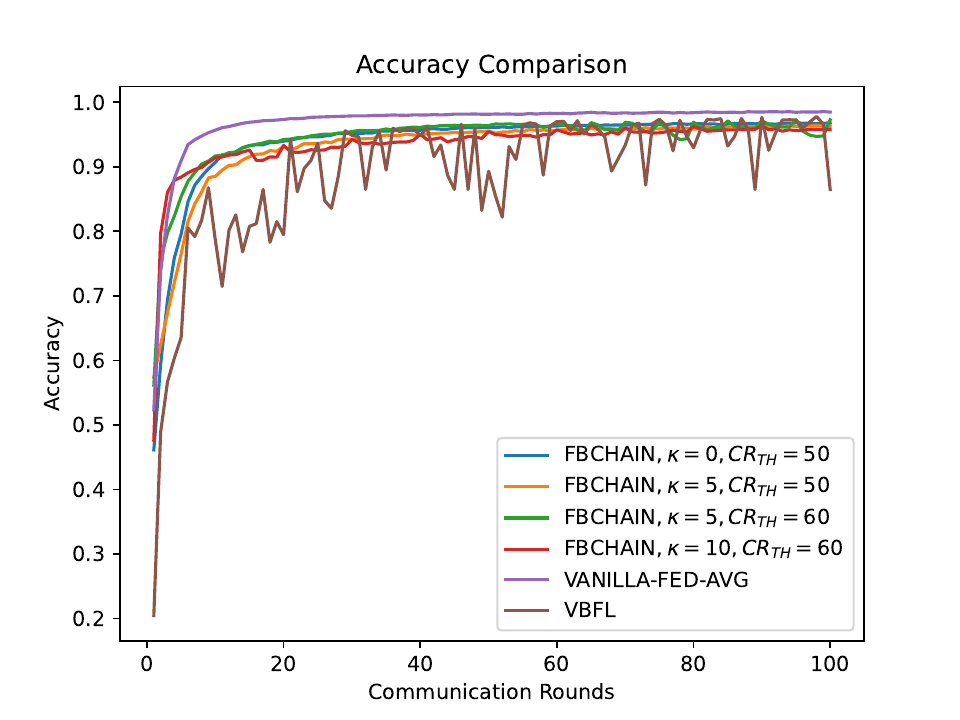}
\caption{Effectiveness of FBChain, VANILLA-FEDAVG, and VBFL in Non-IID Data Distribution}
\label{FBChain Accuracy Compared with VANILLA-FEDAVG, VBFL}
\end{figure} 


Figure \ref{FBChain Accuracy Compared with VANILLA-FEDAVG, VBFL} demonstrates the effectiveness of FBChain, our proposed federated learning model, by showing the global model accuracy trend over 100 training rounds. We compare FBChain with two other models: vanilla federated average learning (VFL), which is shown as \textbf{VANILLA-FED-AVG} in the figure, and VBFL  \cite{Chen2021RobustBF}, which introduces a novel decentralized validation mechanism,
a novel decentralized verification mechanism is introduced, where the legitimacy of local model updates is reviewed by a single verifier. Additionally, they designed a specialized Proof-of-Stake consensus mechanism, in which stakes are more frequently rewarded to reputable devices, thereby safeguarding legitimate local model updates by increasing their chances of dominating blocks on the blockchain. 
 We assign FBChain to 20 devices, including 12 $LT$, 3 $PA$, and 5 $BP$ nodes, and compare it with VFL assigned to 20 clients and VBFL assigned to 12 workers, 5 validators, and 3 miners. We use different values of $\kappa$ and $CR_{TH}$ for FBChain, with default values of $CR^r$ and $CR^p$ set to 5 and -5, respectively. When $\kappa = 10$ and $CR_{TH} = 60$, $CR^r$ and $CR^p$ are adjusted to 10 and -10, respectively. VBFL is assigned a validator-threshold of 0.08 and no malicious nodes. The purple and brown curves represent the global model accuracy trend for VFL and VBFL, respectively, while the other curves show the performance of FBChain with different values of $\kappa$, $CR_{TH}$, $CR^r$, and $CR^p$. When $\kappa = 0$, all $LT$'s local models participate in the global model update.
When $\kappa = 5$, only $LT$ nodes with a $CR$ value greater than or equal to $CR_{TH}$ are allowed to participate in the global model update every round. If a $LT$ node has a $CR$ value less than $CR_{TH}$, it can only participate in the update every 5 rounds, and when $\kappa = 10$, the round number is increased to 10. Despite having a relatively small number of $LT$ nodes, FBChain maintains a high level of accuracy compared to the vanilla federated learning model.


\subsection{Transmission Delay}

\begin{figure*}[htbp]
\centering
\begin{subfigure}[b]{0.3\textwidth}
	\centering
	\includegraphics[width=\textwidth]{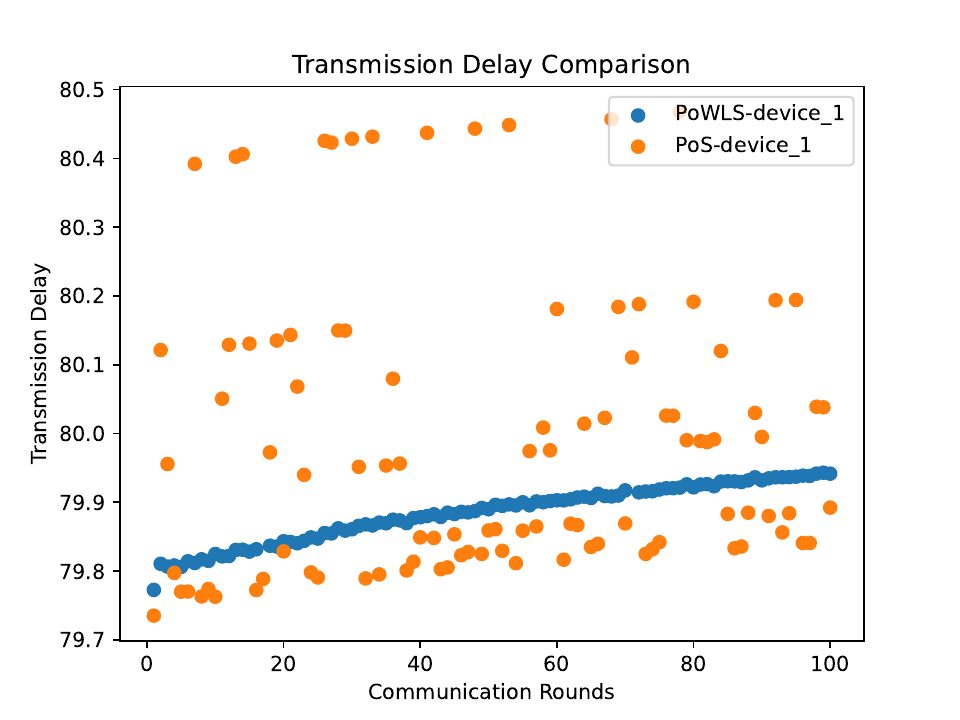}
	\caption{Device 1}
\end{subfigure}
\begin{subfigure}[b]{0.3\textwidth}
	\centering
	\includegraphics[width=\textwidth]{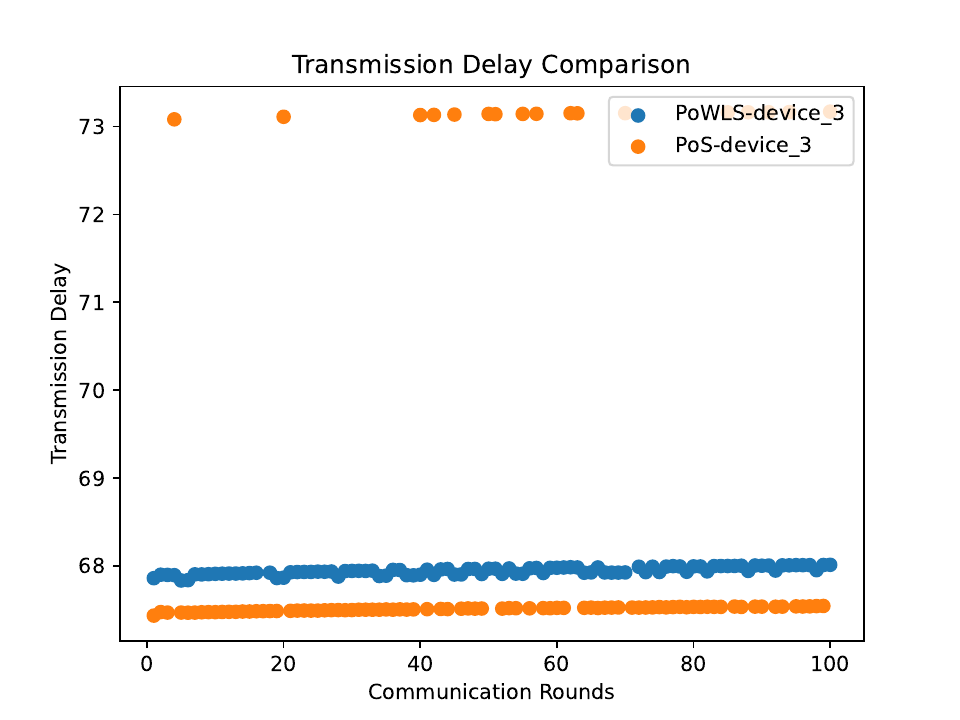}
	\caption{Device 3}
	\label{Device 3}
\end{subfigure}
\begin{subfigure}[b]{0.3\textwidth}
	\centering
	\includegraphics[width=\textwidth]{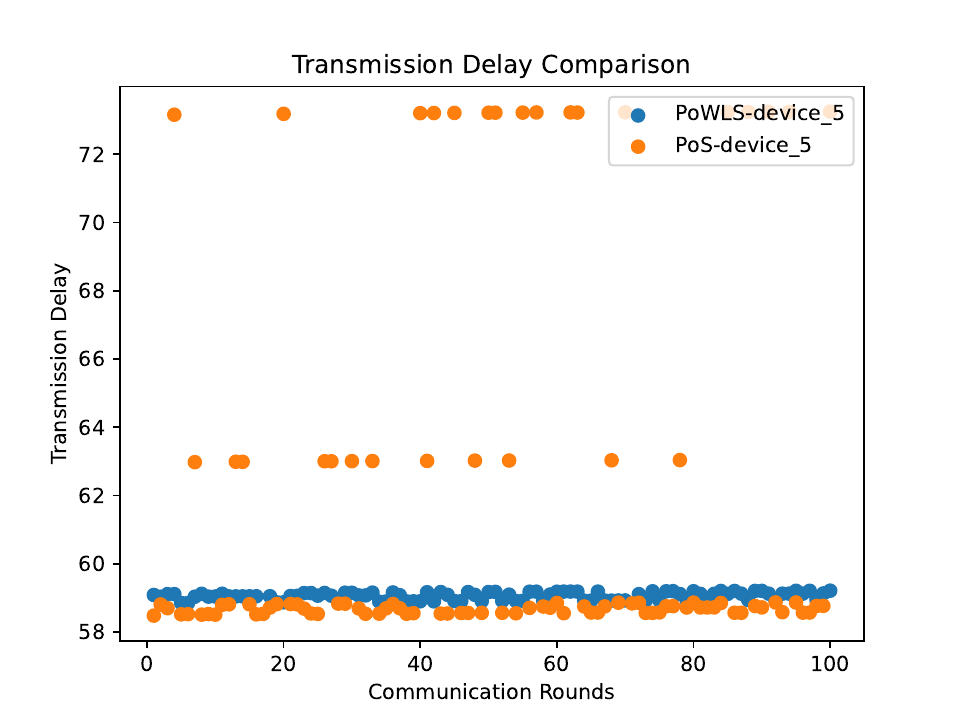}
	\caption{Device 5}
	\label{Device 5}
\end{subfigure}

\begin{subfigure}[b]{0.3\textwidth}
	\centering
	\includegraphics[width=\textwidth]{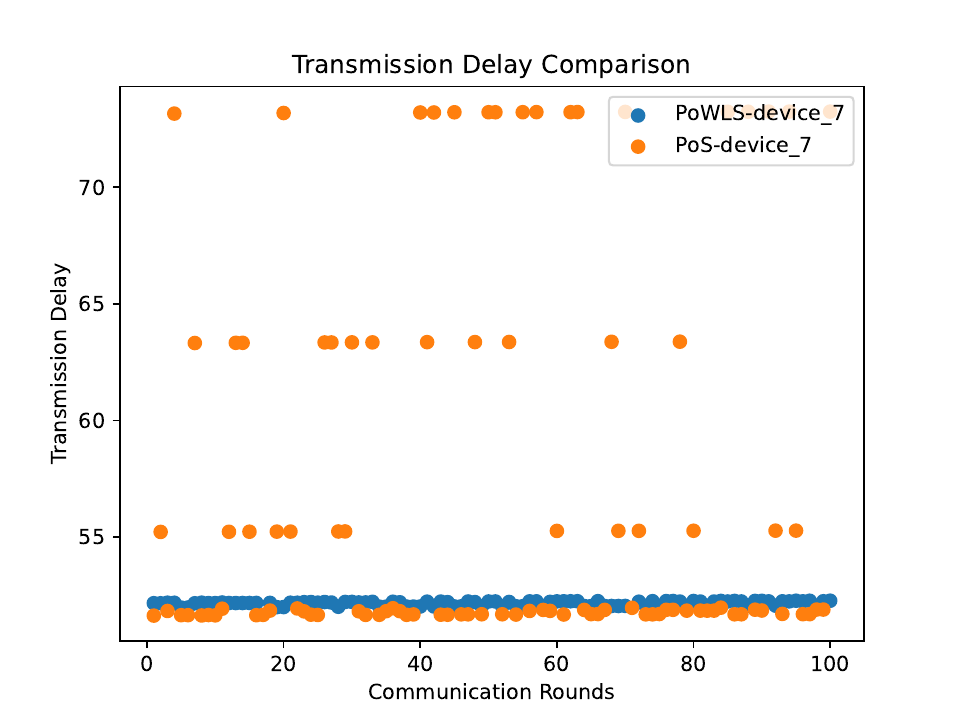}
	\caption{Device 7}
	\label{Device 7}
\end{subfigure}
\begin{subfigure}[b]{0.3\textwidth}
	\centering
	\includegraphics[width=\textwidth]{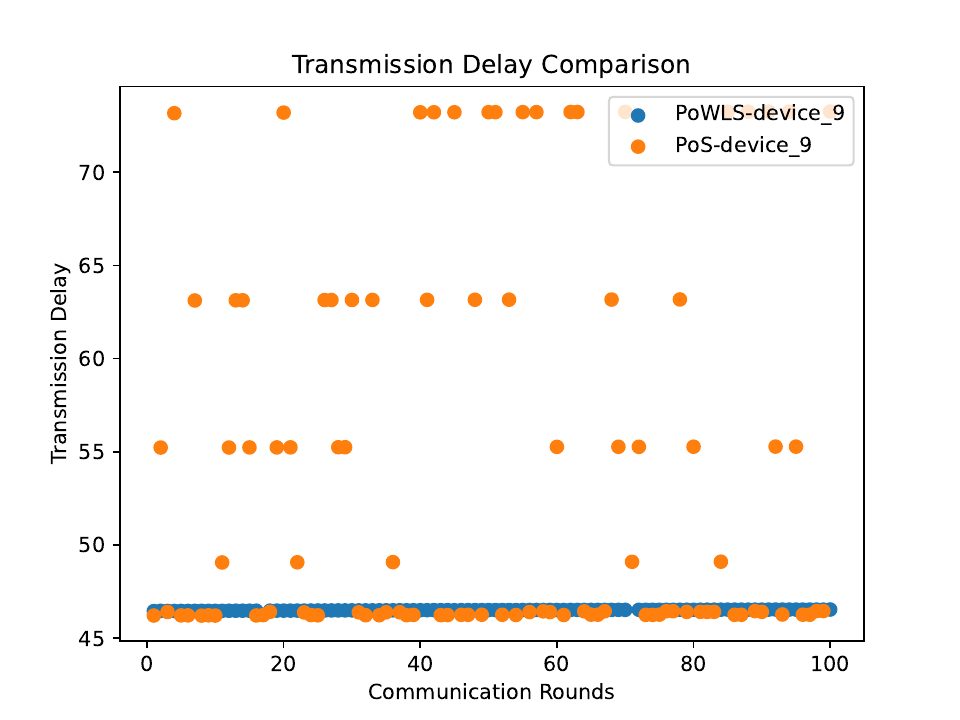}
	\caption{Device 9}
	\label{Device 9}
\end{subfigure}
\begin{subfigure}[b]{0.3\textwidth}
	\centering
	\includegraphics[width=\textwidth]{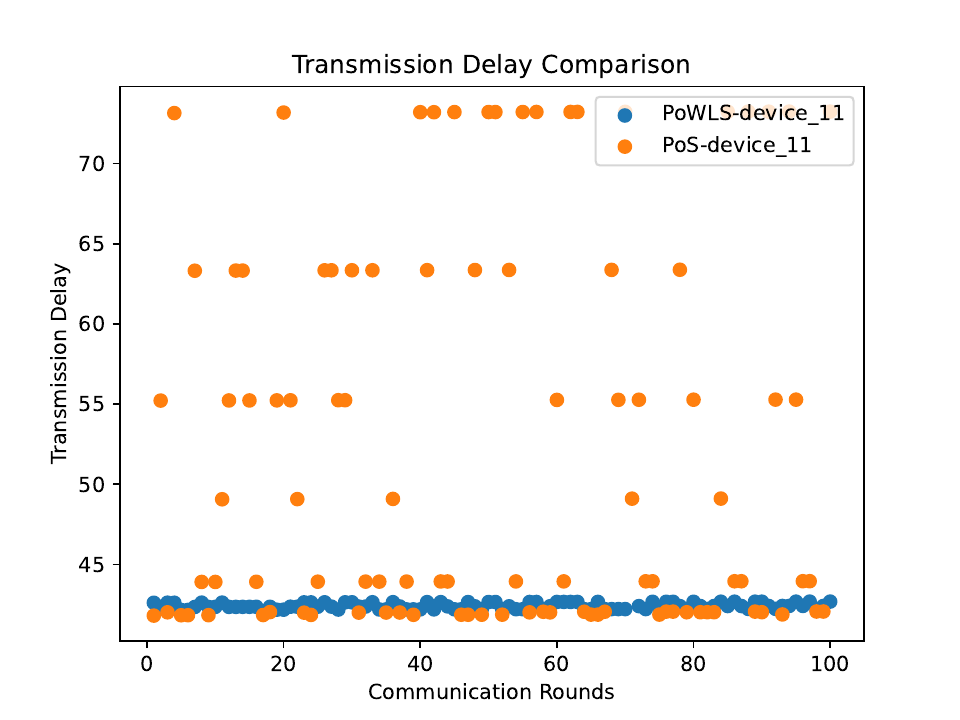}
	\caption{Device 11}
	\label{Device 11}
\end{subfigure}

\begin{subfigure}[b]{0.3\textwidth}
	\centering
	\includegraphics[width=\textwidth]{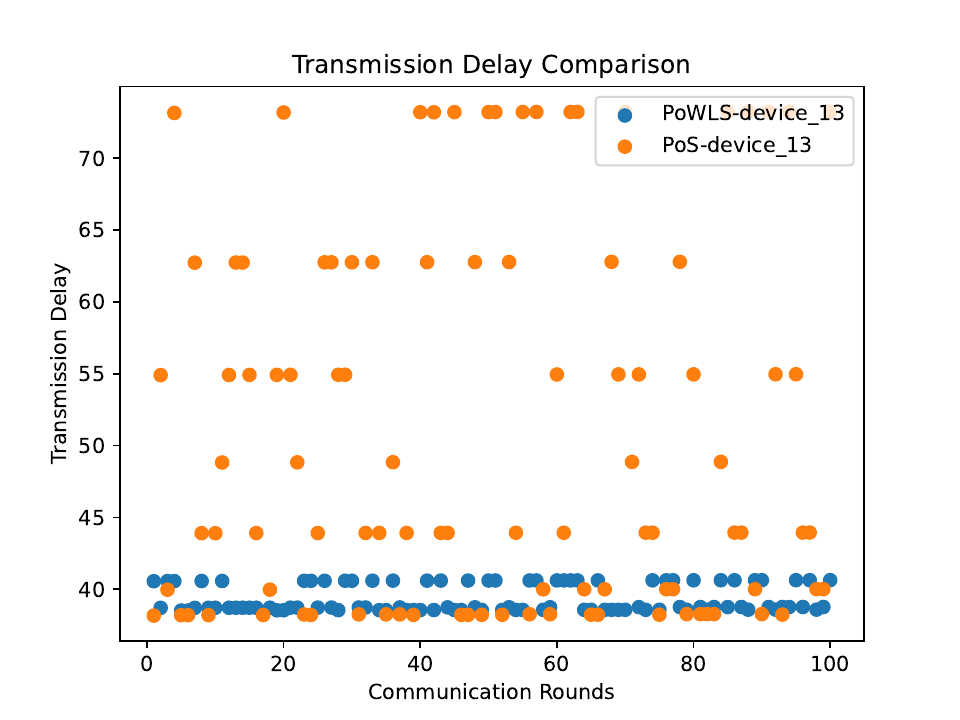}
	\caption{Device 13}
	\label{Device 13}
\end{subfigure}
\begin{subfigure}[b]{0.3\textwidth}
	\centering
	\includegraphics[width=\textwidth]{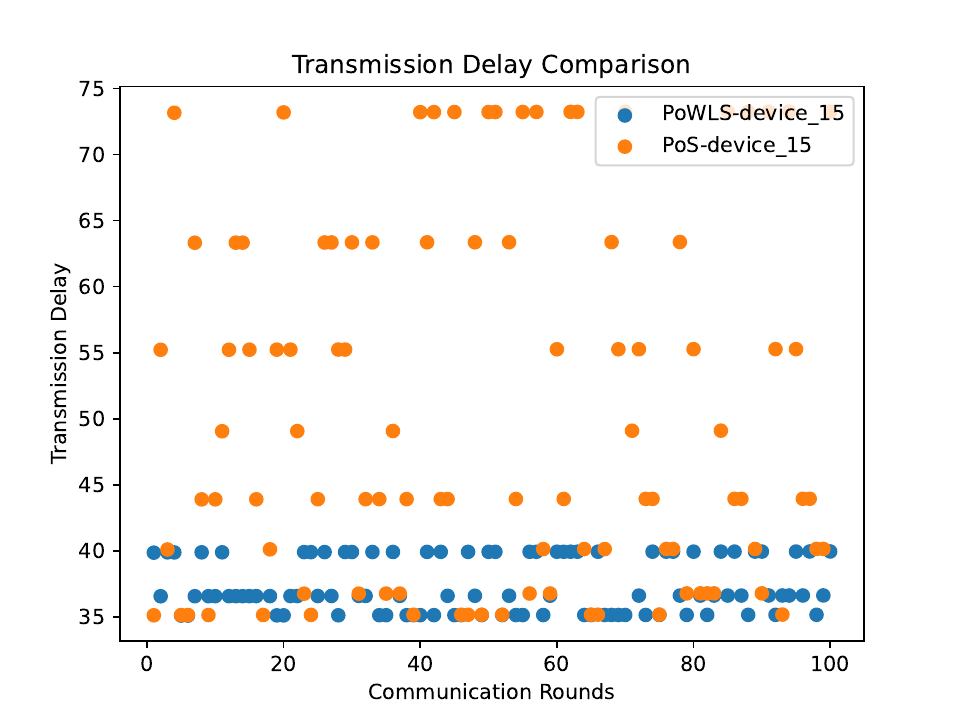}
	\caption{Device 15}
	\label{Device 15}
\end{subfigure}
\begin{subfigure}[b]{0.3\textwidth}
	\centering
	\includegraphics[width=\textwidth]{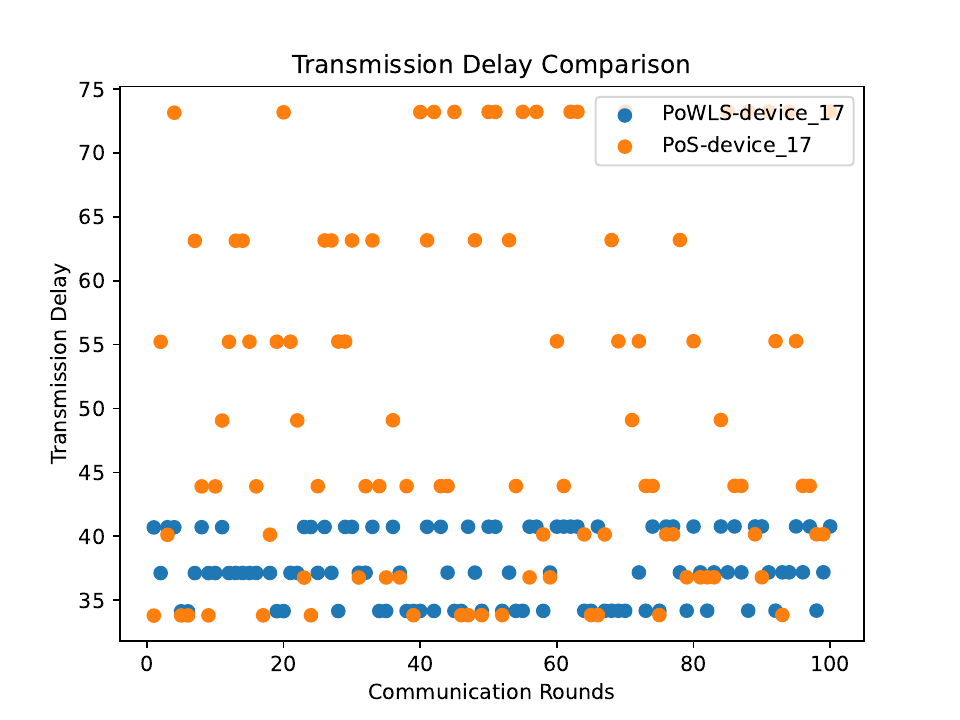}
	\caption{Device 17}
	\label{Device 17}
\end{subfigure}

\begin{subfigure}[b]{0.3\textwidth}
	\centering
	\includegraphics[width=\textwidth]{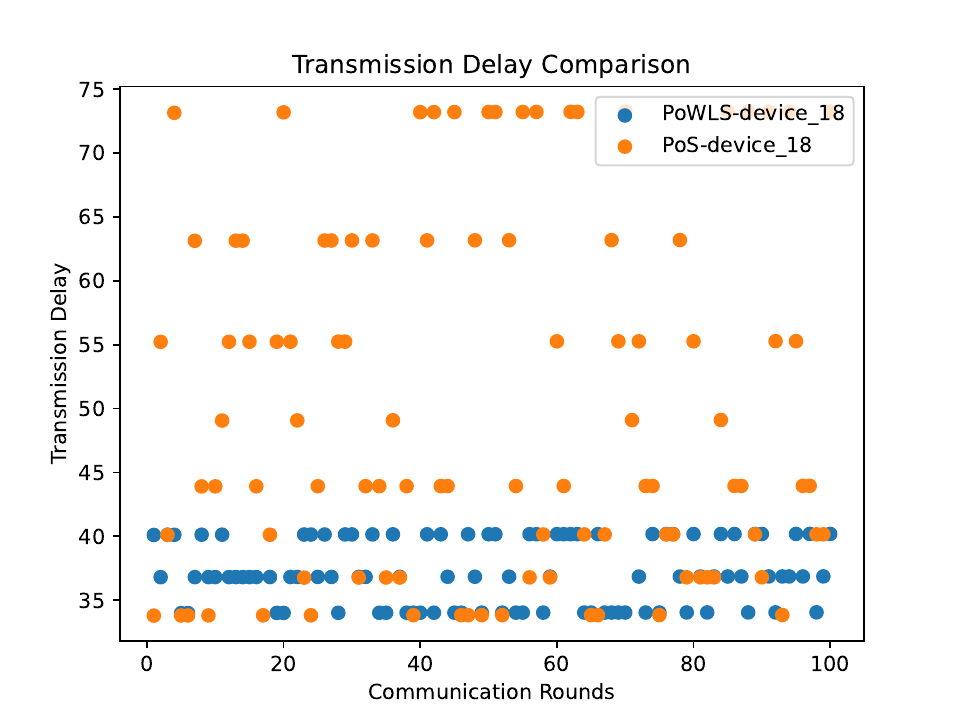}
	\caption{Device 18}
	\label{Device 18}
\end{subfigure}
\begin{subfigure}[b]{0.3\textwidth}
	\centering
	\includegraphics[width=\textwidth]{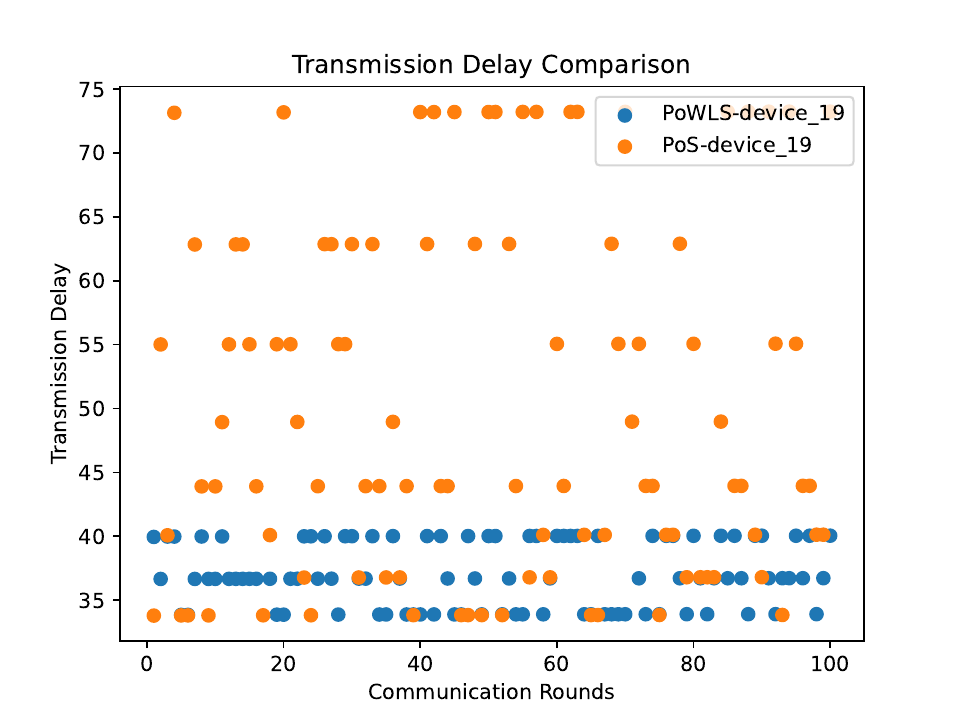}
	\caption{Device 19}
	\label{Device 19}
\end{subfigure}
\begin{subfigure}[b]{0.3\textwidth}
	\centering
	\includegraphics[width=\textwidth]{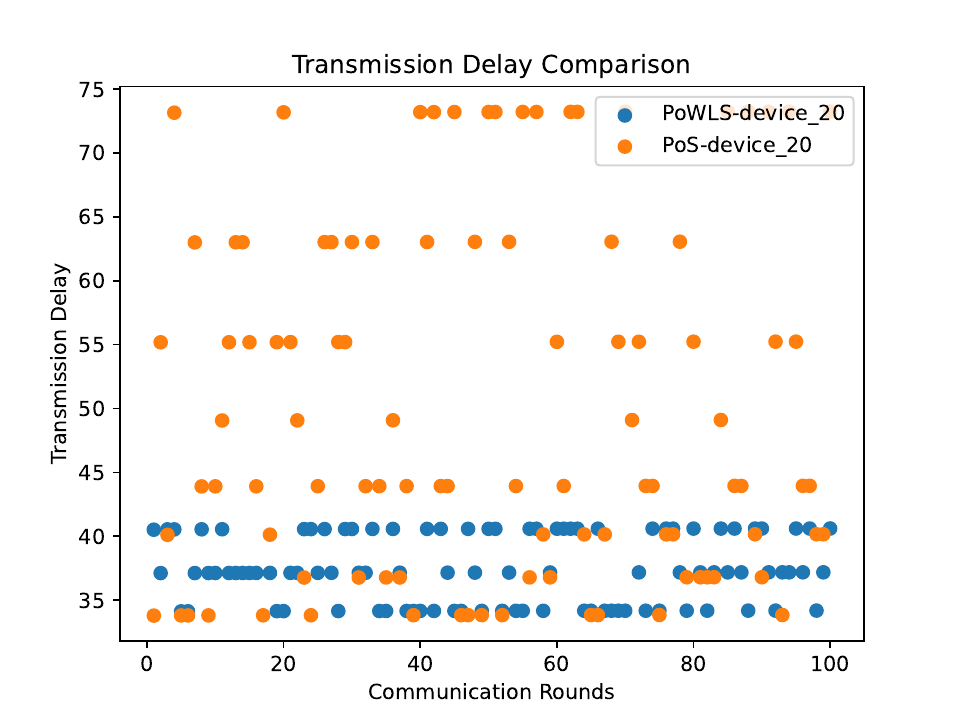}
	\caption{Device 20}
	\label{Device 20}
\end{subfigure}
\caption{Transmission Delay of Local Training Devices, $\kappa = 0, CR_{TH} = 50, CR^r = 5,CR^p = -5$}
\label{Transmission Delay of Local Training Devices}
\end{figure*}

Figure \ref{Transmission Delay of Local Training Devices} shows the transmission delay between $LT$ and $PA$ in PoWLS and PoS of FBChain, we assigned $LT_{\zeta}, \zeta \in [1,20]$, link speed increases from 70000 bytes/s with the increase of $\zeta$, $D_{\zeta} = 70000 + 7000 \times \zeta$, select nodes with evenly distributed link speed from all nodes as $LT$, selected $PA$ from remaining nodes by PoWLS. To demonstrate the effectiveness of PoWLS, we designate the $LT$ as  Device 1,3,5,7,9,11,13,15,17,18,19,20 in PoWLS.
$TD_{\zeta}$ is randomly assigned in [0,1] seconds. From Figure \ref{Transmission Delay of Local Training Devices}, we can find with the device link speed increases, 
the transmission time of the PoS consensus algorithm is gradually greater in more rounds compared to PoWLS due to the different link speeds of $PA$, the transmission speed between $LT$ and $PA$  is constrained by the lower speed nodes, resulting in differences in transmission time among different rounds. For different $LT$, when $D_{LT}<D_{PA} $, the maximum transmission speed between $LT $and $PA $is $D_{LT}$, therefore in Device 1, because $D_1 < D_{\digamma}, 1 < \digamma \leq 20 $, so no matter consensus algorithm is PoWLS or PoS, the transmission time is stable between [79.7, 80.5]. In \ref{Device 3}, because only Device 2 has a link speed lower than Device 3 when Device 2 is $PA$,  the transmission speed will be limited by Device 2, and transmission time will be higher than transmission to other $PA$. It can be seen that in Figure  \ref{Device 3}, the transmission delay of some communication rounds is higher than that of other rounds in PoS. In Figure  \ref {Device 3}, PoWLS and PoS are consistent for most of the time, but in most cases, PoWLS is slightly higher than PoS due to differences in the amount of data transmitted. From Device 5, we can see that in more rounds, the transmission delay of PoS is higher than that of PoWLS, and the transmission delay distribution of Device 17, 18, 19, and 20 tends to be consistent because $PA $nodes are selected from nodes other than $LT$, the $PA$ has highest link speed is Device 16, $D_{16} < D_{17} < D_{18} < D_{19} < D_{20}$, the maximum transmission speed limited by Device 16, so the transmission delay is similar. 

The transmission delay between $LT$ and $PA$ can be seen in Figure \ref{Transmission Delay of Local Training Devices} that in PoS, the transmission delay is unstable and high, while the transmission delay of PoWLS is kept in a low range because PoWLS choose $PA$ by link speed and latency, nodes with faster link speed and lower latency will become $PA$, but in PoS the witness node will be chosen based on the number of stakes.


\subsection{Credit Score and Stake Trending}

\begin{figure}[htbp]
\centering
\begin{subfigure}[b]{0.4\textwidth}
	\centering
	\includegraphics[width=\textwidth]{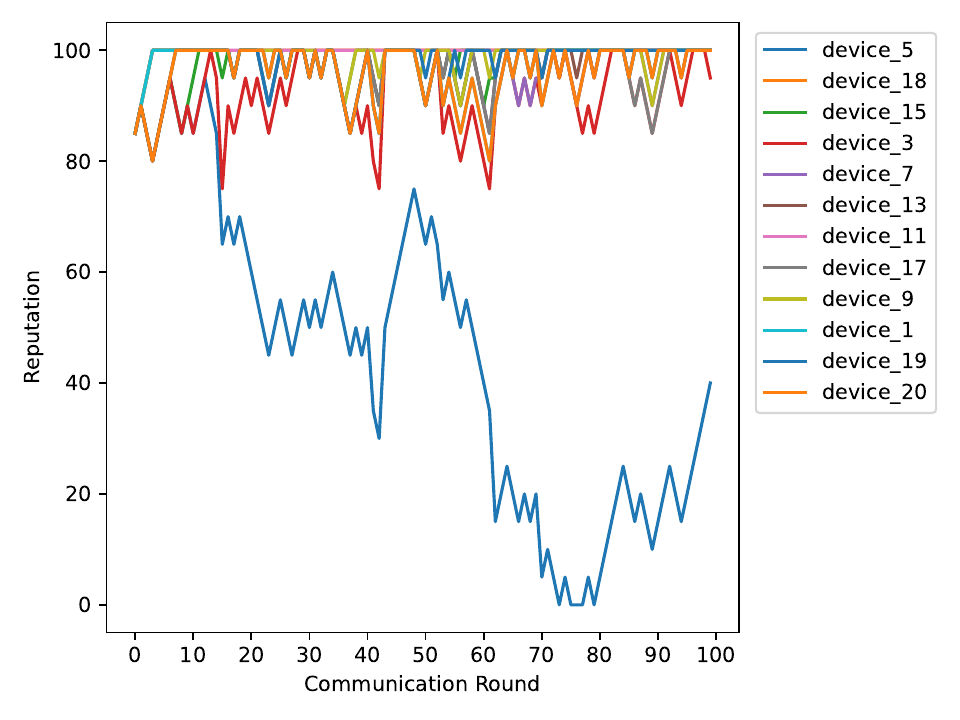}
	\caption{$\kappa = 0, CR_{TH} = 50, CR^r = 5,CR^p = -5$}
	\label{kappa0}
\end{subfigure}

\begin{subfigure}[b]{0.4\textwidth}
	\centering
	\includegraphics[width=\textwidth]{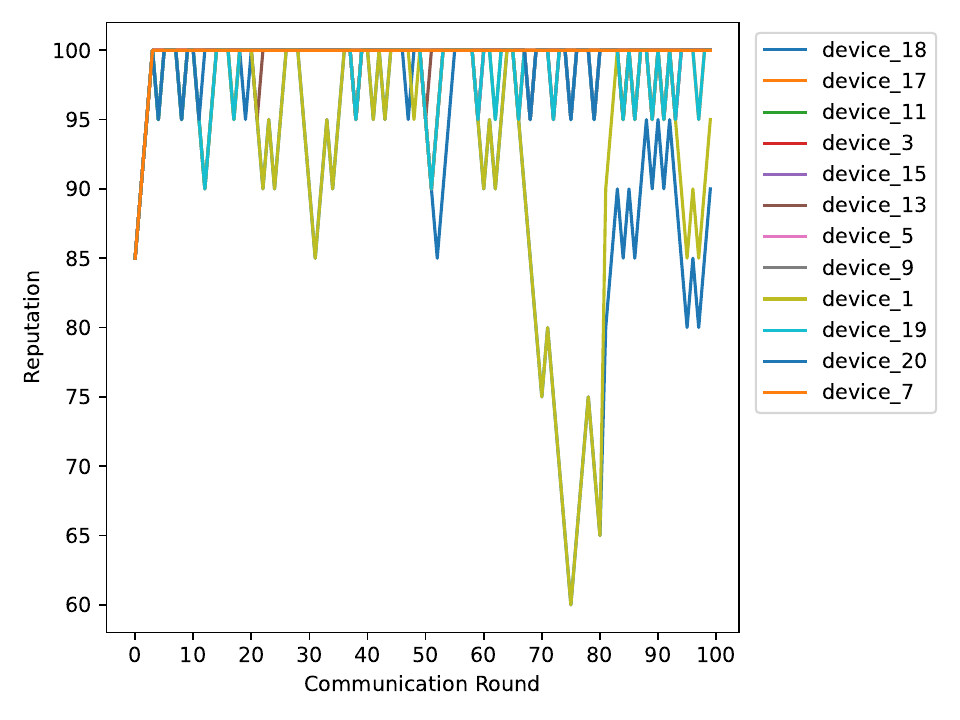}
	\caption{$\kappa = 5, CR_{TH} = 50, CR^r = 5,CR^p = -5$}
	\label{kappa5}
\end{subfigure}

\begin{subfigure}[b]{0.4\textwidth}
	\centering
	\includegraphics[width=\textwidth]{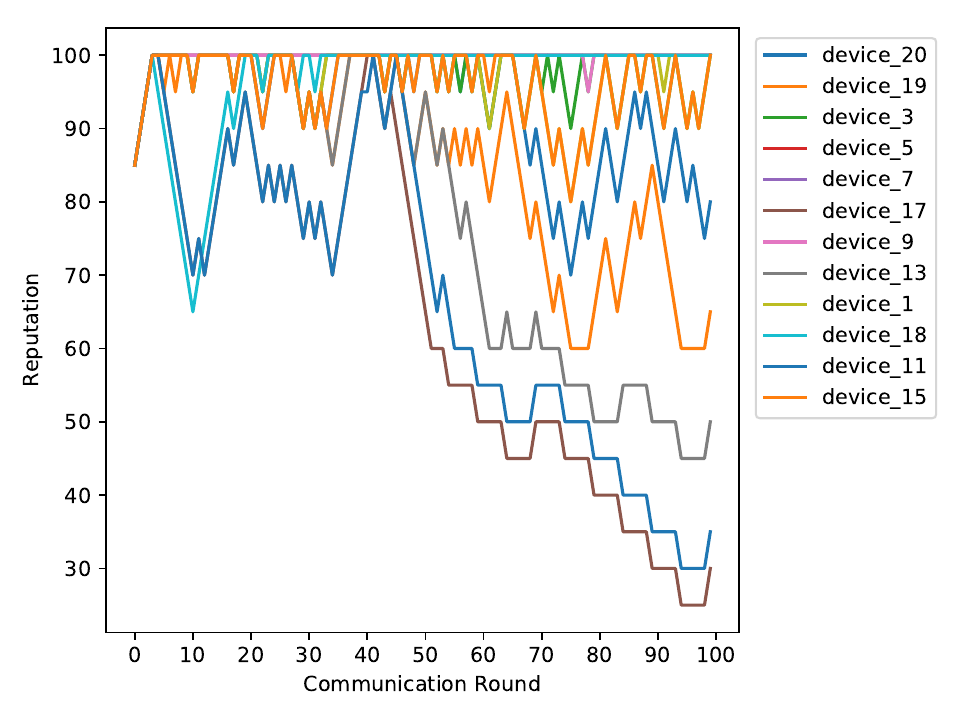}
	\caption{$\kappa = 5, CR_{TH} = 60, CR^r = 5,CR^p = -5$}
	\label{kappa56}
\end{subfigure}

\begin{subfigure}[b]{0.4\textwidth}
	\centering
	\includegraphics[width=\textwidth]{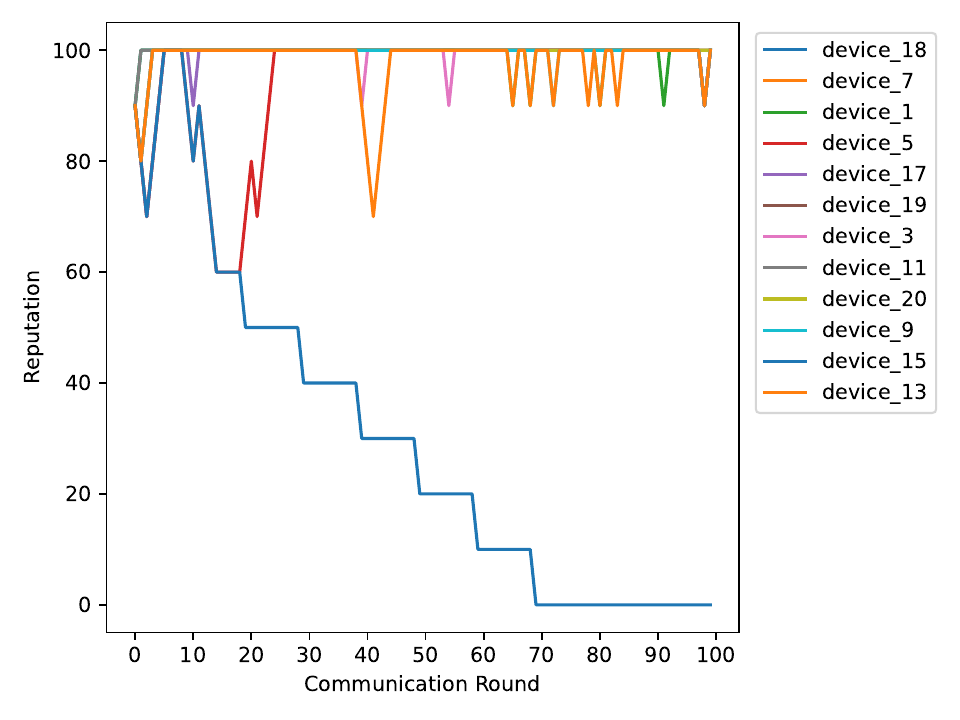}
	\caption{$\kappa = 10, CR_{TH} = 60, CR^r = 10,CR^p = -10$}
	\label{kappa10}
\end{subfigure}
\caption{Local Training Devices Credit Score Trends}
\label{Local Training Devices credit Score Trends}
\end{figure}

\begin{figure}[htbp]
\centering
\includegraphics[width=.5\textwidth]{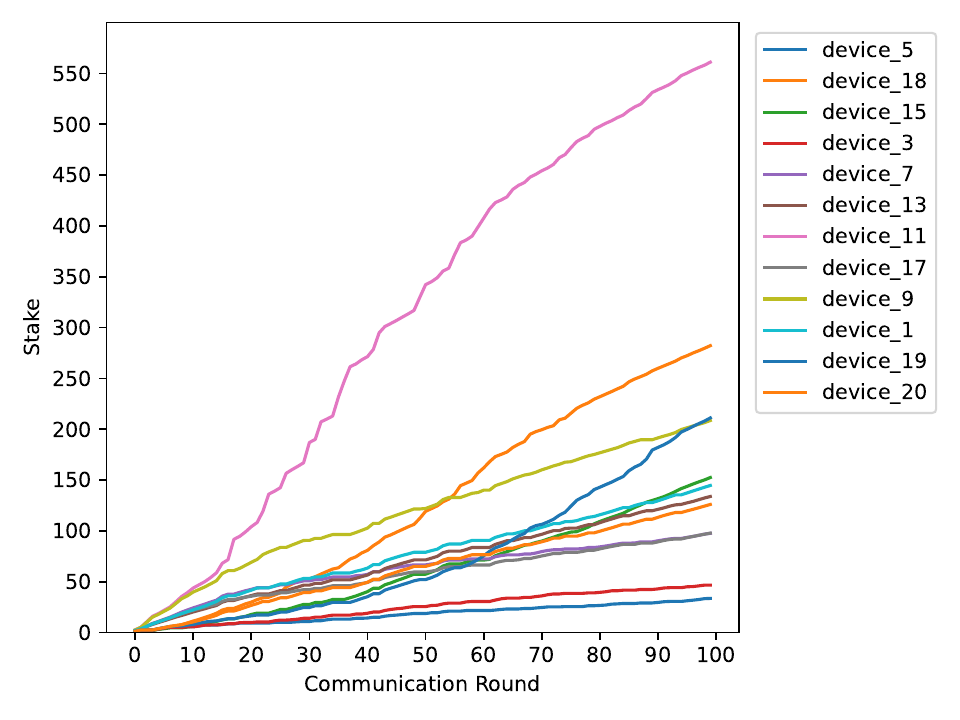}
\caption{Local Training Devices Stake Trends,$\kappa = 0, CR_{TH} = 50, CR^r = 5,CR^p = -5$}
\label{Local Training Devices Stake Trends}
\end{figure}

In Figure \ref{Local Training Devices credit Score Trends} and Figure \ref{Local Training Devices Stake Trends} shows the $CR$ trends of FBChain for $\kappa = \{0,5,10\} , CR_{TH} = \{50,60\}, CR^r = \{5, 10\}, CR^p = \{-5,-10\}$, and assigned $CR \in [0,100]$. In Figure \ref{kappa0} shows while assigning FBChain $\kappa = 0, CR_{TH} = 50, CR^r = 5, CR^p = -5$ the $CR$ trends, we can find the $LT$ besides device\_5 are maintained a high $CR$, and device\_5's $CR$ flowed to 0, after 80 communication round it has an increase, that because when testing device\_5's $LM$ on the test set, the accuracy difference between the results obtained and the global model is lower than $t_{acc}$, so the deduction is made to the $CR$ of device\_5, and when it is larger than $t_{acc}$, benefit device\_5 with $CR$. In Figure \ref{Local Training Devices Stake Trends} shows the stake trend of $LT$,  we assigned $TR_{total} = 20$, we can find device\_11  grows fast, which means it has better performance rather than other $LT$, and in every round, $LT$ distribute rewards from $TR_{total}$ based on the accuracy of $LT$'s $LM$ accuracy performance.In Figure \ref{kappa5} shows  while assign FBChain $\kappa = 5, CR_{TH} = 50, CR^r = 5,CR^p = -5$ the $CR$ trends, and there has no $LT$'s $CR < CR_{TH}$. Every $LT$ participates in the update of $GM$ in every round. In Figure \ref{kappa56} shows  while assign FBChain $\kappa = 5, CR_{TH} = 60, CR^r = 5,CR^p = -5$ the $LT$'s $CR$ trends, and we can find the $CR$ of three nodes has been less than $CR_{TH}$ for a period of time, respectively device\_17,device\_11 and device\_13. After their $CR$ lower than $CR_{TH} = 60$, they can only participate in the update of $GM$ in every $\kappa = 5 $ rounds, we can find their $CR$ also changes in every 5 rounds when their $CR < 60$. In Figure \ref{kappa10} shows  while assign FBChain $\kappa = 10, CR_{TH} = 60, CR^r = 10,CR^p = -10$ the $LT$'s $CR$ trends, from Figure \ref{kappa10} we can find device\_15's $CR$ reached lower than $CR_{TH} = 60$ in first 20 communication rounds, then it changed every $\kappa = 10$ rounds, which is the device\_15 participate in the update of $GM$ in every $\kappa = 10 $ rounds.

\section{Conclusion and Future Work}
\label{Conclusion}

In this paper, we propose FBChain, a federated learning blockchain model that improves communication efficiency while preventing potential data tampers and leakage during model parameter transmission and reducing blockchain storage pressure. The PoWLS consensus algorithm introduced by FBChain selects nodes with better network link speed and latency for global model aggregation and block package, thereby improving the efficiency of data transmission between local training nodes and aggregation nodes. Our focus in this paper is on improving the communication efficiency and security of federated learning, and we have provided validation for this approach. However, further research is needed to address the issue of training resource utilization and imbalanced training data.

\end{document}